\documentclass[twocolumn,preprintnumbers,amsmath,amssymb]{revtex4}
\usepackage{graphicx,epsfig}
\usepackage{dcolumn}
\usepackage{bm}
\newpage

\begin{document}

\title{Frequency dependent heat capacity within a kinetic model of 
glassy dynamics}
\author{Dwaipayan Chakrabarti and Biman Bagchi\footnote[2] 
{For correspondence: bbagchi@sscu.iisc.ernet.in}}
\affiliation{Solid State and Structural Chemistry Unit, Indian Institute of 
Science, Bangalore 560012, India}

\date \today

\begin{abstract}

There has been renewed interest in the frequency dependent specific heat 
of supercooled liquids in recent years with computer simulation 
studies exploring the whole frequency range of relaxation. The simulation
studies can thus supplement the existing experimental results to provide an 
insight into the energy landscape dynamics. We here investigate a kinetic 
model of cooperative dynamics within the landscape paradigm for the dynamic 
heat capacity $C(\omega, T)$ behavior. In this picture, the $\beta$-process 
is modeled as a thermally activated event in a two-level system and the 
$\alpha$-process is described as a $\beta$-relaxation mediated cooperative 
transition in a double well. The model resembles a landscape picture, 
apparently first conceived by Stillinger [Science {\bf 267}, 1935 (1995)], 
where an $\alpha$-process is assumed to involve a concerted series of 
$\beta$-processes. The model provides a description of the activated hopping
in the energy landscape in close relation with the cooperative nature of the
hopping event. For suitable choice of parameters, the model predicts a 
frequency dependent heat capacity that reflects the two-step relaxation 
behavior. Although experimentally obtained specific heat spectra of 
supercooled liquids till date could not capture the two-step relaxation 
behavior, this has been observed in a computer simulation study by Scheidler
{\it et. al.} [Phys. Rev. B {\bf 63}, 104204 (2001)]. The separation between
the two peaks grows as the temperature drops, indicating the stringent 
constraint on the $\alpha$-process due to the cooperativity requirement. The temperature dependence of the position of the low-frequency peak, due to the 
$\alpha$-relaxation, shows a non-Arrhenius behavior as observed 
experimentally by Birge and Nagel [Phys. Rev. Lett. {\bf 54}, 2674 (1985)]. 
The {\it shape} of the $\alpha$-peak is, however, found to be temperature 
independent, in agreement with the simulation result. The high-frequency 
peak appears with considerably larger amplitude than the $\alpha$-peak.
We attempt a plausible reason for this observation that is in contrast 
with the general feature revealed by the dielectric spectroscopy. The
relative amplitudes of the $\beta$- and $\alpha$-peaks in the present 
framework are found to depend on several characteristic features of the 
energy landscape, including the extent of cooperativity requirement for the 
$\alpha$-relaxation and the asymmetry of the double well.

\end{abstract}

\newpage 

\maketitle

\noindent{\bf I. INTRODUCTION}

Understanding the complex relaxation phenomena in supercooled liquids 
has motivated much scientific efforts over decades 
\cite{Angell-Ngai-McKenna-McMillan-Martin, 
Debenedetti-Stillinger-Nature-2001, Mohanty-ACP-1994}. The measurement of 
frequency dependent specific heat, pioneered independently by Birge and 
Nagel and by Christensen in the year 1985 
\cite{Birge-Nagel-PRL, Christensen}, opened up another approach to this 
goal. Zwanzig subsequently showed on the basis of linearized hydrodynamics 
that $c_{p}(\omega)$, the frequency dependent specific heat at constant 
pressure, could be directly related to the frequency dependent longitudinal 
viscosity $\eta_{l}(\omega)$ \cite{Zwanzig-JCP-1988}. The frequency 
dependent specific heat has, however, continued to get explored from both 
experimental \cite{Birge-PRB-1986, Dixon-Nagel-PRL-1988, 
Menon-JCP-1997} and theoretical \cite{Oxtoby-JCP-1986, Jackle-PhysicaA, 
Nielsen-Dyre-PRB, Nielsen-PRE-1999, Scheidler-PRB-2001} perspectives in the
hope that specific heat spectroscopy would provide an insight into the 
energy landscape dynamics. 
 
In practice, one measures the frequency dependent specific heat in the
linear response regime following an arbitrary small thermal perturbation 
that takes the system slightly away from the 
equilibrium \cite{Birge-Nagel-PRL}. $c_{p}(\omega)$ is a linear 
susceptibility describing the response of the system to this 
perturbation. One can, however, calculate the frequency dependent 
specific heat in terms of equilibrium fluctuation of energy following a 
relevant fluctuation-dissipation theorem derived explicitly by Nielsen and 
Dyre for a system whose dynamics is described by a master equation 
\cite{Nielsen-Dyre-PRB}. In this work, we do so for a kinetic 
model of glassy dynamics that invokes the concept 
$\beta$-organized-$\alpha$-process \cite{Chakrabarti-Bagchi-JCP} 
within the landscape paradigm. 

The measurements of frequency dependent specific heat currently suffer 
from a limitation that frequency range up to $10^4$ Hz can only be 
probed with the presently available experimental setup. Therefore, the 
experimentally obtained frequency spectra capture only the 
$\alpha$-relaxation regime of supercooled liquids. However, Sceidler 
{\it et. al.} have recently carried out a computer simulation study of a 
system that models amorphous silica, where they could scan the whole 
frequency range of interest, revealing the two-peak structure with a notably {\it dominant} high-frequency peak \cite{Scheidler-PRB-2001}. The 
high-frequency peak shows only a weak temperature dependence and has been 
ascribed to the vibrational excitations of the system. In their work 
\cite{Scheidler-PRB-2001}, they have applied the Mori-Zwanzig projection 
operator formalism and made use of an exact transformation formula, due to 
Lebowitz {\it et. al.} \cite{Lebowitz-Percus-Verlet}, to derive a relation
between the frequency dependent specific heat $c_{v}(\omega)$ and the 
autocorrelation function of the temperature fluctuations in the 
microcanonical ensemble. This relationship, which is identical to the 
one derived independently by Nielsen in terms of a fluctuation-dissipation 
theorem, has allowed the determination of $c_{v}(\omega)$ from computer 
simulations in equilibrium \cite{Nielsen-PRE-1999}. A mode coupling theory 
(MCT) based calculation has also shown the two-step relaxation behavior in 
the predicted frequency spectrum of the specific heat 
\cite{Harbola-Das-PRE}. The dominance of the high-frequency peak is, 
however, not evident in this work. 

The measurements of frequency dependent specific heat in specific heat
spectroscopy supplemented by computer simulation studies may prove to be 
useful in providing insight into the landscape dynamics of supercooled 
liquids. The landscape paradigm has been widely used to elucidate dynamics 
of liquids in the supercooled regime \cite{Goldstein-JCP-1969, 
Johari-Goldstein, Stillinger-Science-1995, Sastry-Nature-1998}. 
This framework involves the division of the multidimensional configuration 
space into so called {\it metabasins} on the basis of a transition 
free-energy criterion. Two vastly different timescales thus get entailed, 
the smaller one due to motions {\it within} the metabasins and the longer 
one due to exchange {\it between} the metabasins involving much larger 
free-energy of activation. In particular, the $\beta$-processes are 
visualized to originate from {\it activated dynamics within a metabasin}, 
while {\it escape from one metabasin to another} is taken to describe an 
$\alpha$-process \cite{Stillinger-Science-1995}. See {\bf Figure 1} for 
a schematic representation of the two processes. It is important to 
note here that the breakdown of the MCT \cite{Bengtzelius, Gotze-Sjogren, 
Angell-JPCS} is ascribed to the dominance of relaxation by these thermally 
activated hopping events \cite{Bagchi-JCP-1994}, which are unaccounted for 
in the ideal version of MCT. Recent computer simulation studies have further revealed that hopping is a highly cooperative phenomenon promoted by many 
body fluctuations \cite{hopping1, hopping2, hopping3, hopping4}; hopping of 
a tagged particle is often preceded by somewhat larger than normal, but 
still small amplitude motion of several of its neighbors \cite{hopping4}. A 
rather different stringlike cooperative motion has also been found to occur 
in a model glass-forming liquid \cite{Donati-et-al-PRL-1998}. In a very 
recent simulation study that investigates the role of cooperativity in
reorientational and structural relaxation in a supercooled liquid both on 
the potential energy landscape and in the real space, it has been further
proposed that relaxation involves activation of the system to the complex 
multidimensional region on the borders of the basins of attraction of the 
minima for an extended period \cite{Kim-Keyes-JCP-2004}.  
\begin{figure}[tb]
\epsfig{file=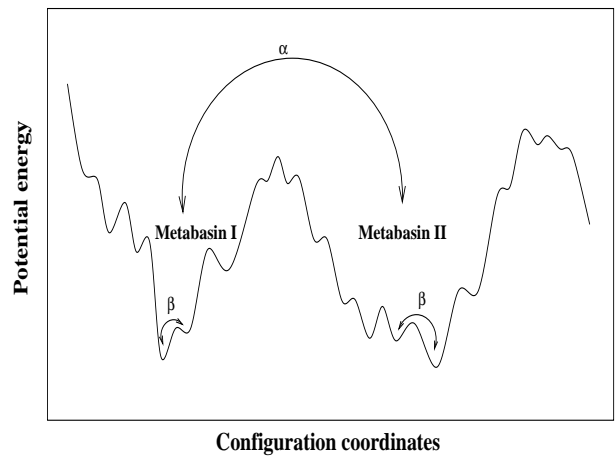,height=6cm,width=8cm,angle=0}
\caption{A schematic representation of the potential energy landscape 
showing motions within and between metabasins. }
\end{figure}

In the present work, we employ a kinetic model of glassy dynamics that 
attempts to provide a description of the activated hopping within the 
landscape paradigm in close connection with the cooperative nature of 
the hopping event. We follow a procedure, as outlined by Nielsen and 
Dyre \cite{Nielsen-Dyre-PRB}, to compute the frequency dependent heat 
capacity $C(\omega,T)$ for our model system. For suitable choice of 
parameters, the model predicts a frequency dependent heat capacity that 
captures many of the features of the two-step relaxation behavior in 
supercooled liquids. The predictions of our model are in good 
qualitative agreement with the available experimental and computer 
simulation results.

The outline of the paper is as follows. In the next section we describe 
the model. Section III provides the theoretical treatment. We present 
the results along with discussion in section IV. Section V concludes 
with a summary of the results and a few comments.   

\noindent{\bf II. DESCRIPTION OF MODEL}

We model a $\beta$-process as an activated event in a two-level system 
(TLS). We label the ground level of a TLS as $0$ and the excited level 
as $1$. The waiting time before a transition can occur from the level $i (= 0, 1)$ is assumed to be random and is given by the Poissonian 
probability density function:
\begin{equation} 
\psi_{i}(t) = \frac{1}{\tau_{i}} exp(-t/\tau_{i}), ~~~~~~~~~~~~~i = 0, 1,
\label{A}
\end{equation}
where $\tau_{i}$ is the average time of stay at the level $i$. If 
$p_{i}(T)$ denotes the canonical equilibrium probability of the level 
$i$ of a TLS being occupied at temperature $T$, the equilibrium constant 
$K(T)$ for the population in two levels at temperature $T$ is given by 
the following relation that obeys the detailed balance:
\begin{equation}
K(T) = \frac{p_{1}(T)}{p_{0}(T)} = \frac{\tau_{1}(T)}{\tau_{0}(T)} = 
exp[-\epsilon/(k_{B}T)],
\label{B}
\end{equation}
where $\epsilon$ is the energy separation between the two levels in a
TLS, and $k_{B}$ is the Boltzmann constant. Here the level $0$ is taken 
to have a zero energy. 

Within the framework of the present model, a metabasin is characterized 
by an $N_{\beta}$ number of such non-interacting two-level systems 
(TLSs). A given minimum number among the total number $N_{\beta}$ of 
TLSs must simultaneously be in the excited levels for the occurrence of 
an $\alpha$-process. We here concentrate on two adjacent metabasins, 
which we label as $1$ and $2$ and together call a double well. 
{\bf Figure 2} shows a schematic diagram of two adjacent metabasins with illustration of dynamics within and between them. The respective numbers of TLSs that comprise the metabasins are $N_{\beta}^{(1)}$ and 
$N_{\beta}^{(2)}$. For a collection of $N_{\beta}^{(i)} ~ (i = 1, 2)$ 
TLSs, a variable $\zeta_{j}^{i}(t),~(j = 1, 2, ....., N_{\beta}^{(i)})$ 
is defined, which takes on a value $0$ if at the given instant of time 
$t$ the level $0$ of the TLS $j$ is occupied and $1$ if otherwise. 
$\zeta_{j}^{i}(t)$ is thus an occupation variable. The collective 
variables $Q_{i}(t)~(i = 1, 2)$ are then defined as
\begin{equation}
Q_{i}(t) = \displaystyle \sum_{j=1}^{N_{\beta}^{(i)}} \zeta_{j}^{i}(t).
\label{C}
\end{equation}
$Q_{i}(t)$ is therefore a stochastic variable in the discrete integer 
space $[0, N_{\beta}^{(i)}]$. $Q_{i}(t)$ serves as an order parameter
for dynamical change involving metabasin {\it i}. Here {\it an 
$\alpha$-process is assumed to occur only when all the $\beta$-processes 
(TLSs) in a metabasin are simultaneously excited, i.e., when 
$Q_{i} = N_{\beta}^{(i)}$.} There is 
a finite rate of transition $k$ from each of the metabasins when this 
condition is satisfied. Within the general framework of the model, the 
double well becomes asymmetric when 
$ N_{\beta}^{(1)} \neq N_{\beta}^{(2)} $, as shown in {\bf Figure 2}.
\begin{figure}[tb]
\epsfig{file=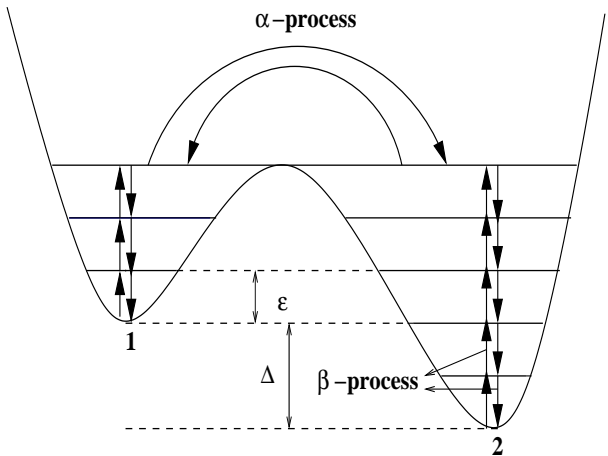,height=6cm,width=8cm,angle=0}
\caption{A schematic representation of the model under 
consideration. The horizontal lines within a well represent different 
excitation levels. Note that the energy levels are in general 
degenerate, as they correspond to the sum of the energies of individual 
TLSs in the collection.}
\end{figure}

It is worth while to note the correspondence of the present description
with real physical processes occurring in glass-formers. The 
$\alpha$-process may correspond to large-scale hopping of a 
particle. For this hopping to occur, however, many small 
reorientations / rearrangements / displacements are required 
simultaneously among its neighbors. The activated dynamics {\it within} 
a TLS may well represent small rotations \cite{Oxtoby-JCP-1986}. In the 
case of polymer melts which exhibit glassy behavior, the 
$\beta$-relaxation may involve the motion of side chains. This picture 
apparently differs from the one drawn by Dyre \cite{Dyre-PRE-1999}, who 
has argued that large-angle rotations are "causes" and small-angle 
rotations are "effects". The present picture, however, contains Dyre's 
one in the sense that small-angle rotations indeed occur following a 
large-scale jump motion for the completion of relaxation as evident in 
{\bf Figure 2}. The present model is built on a rather symmetrical 
picture that also necessitates small-angle rotations for a large-angle 
rotation to occur.  

\noindent{\bf III. THEORETICAL TREATMENT}
 
From a theoretical point of view, the treatment of frequency dependent 
heat capacity can be carried out by employing the linear response 
assumption. Following Nielsen and Dyre \cite{Nielsen-Dyre-PRB}, the 
frequency dependent heat capacity $C(\omega, T)$ of our system at 
temperature $T$ can be given by
\begin{eqnarray}
C(\omega, T) &=& \frac{<E^2(T)>}{k_{B}T^2} \nonumber \\
&-& \frac{s}{k_{B}T^2} 
\int\limits_{0}^{\infty} dt~e^{-st}~<E(t,T)E(0,T)>,
\label{D}
\end{eqnarray}
where $s = i\omega$, $\omega$ being the frequency of the small 
oscillating perturbation, $i = \sqrt{-1}$, and the angular brackets 
denote an equilibrium ensemble averaging. $E(t,T)$ stands for the total 
energy of the system at time $t$ and temperature $T$ and is given by  
\begin{eqnarray}
E(t,T) &=& \displaystyle \sum_{n=0}^{N_{\beta}^{(1)}} P_{1}(n;t,T)
~(N_{\beta}^{(2)} - N_{\beta}^{(1)} + n )~\epsilon \nonumber \\ 
&+& 
\displaystyle \sum_{n=0}^{N_{\beta}^{(2)}} P_{2}(n;t,T)~n~\epsilon.
\label{E}
\end{eqnarray}
Here the lowest level of the well $2$ is taken to have zero energy and 
$P_{i}(n;t,T)$ denotes the probability that the stochastic variable 
$Q_{i}$ takes on a value $n$ in the $i$-th well at time $t$ and
temperature $T$. The evolution of these probabilities obeys the master 
equation \cite{van Kampen}:
\begin{eqnarray}
\frac {\partial P_{i}(n;t,T)}{\partial t} 
&=& [(N_{\beta}^{(i)} - n + 1)/\tau_{0}(T)]P_{i}(n - 1;t,T) \nonumber \\ 
&+& [(n + 1)/\tau_{1}(T)]P_{i}(n + 1;t,T) \nonumber \\ 
&-& [(N_{\beta}^{(i)} - n)/\tau_{0}(T)]P_{i}(n;t,T) \nonumber \\ 
&-& (n/\tau_{1}(T))P_{i}(n;t,T) - k~\delta_{n,N_{\beta}^{(i)}}
~P_{i}(n;t,T) \nonumber \\ 
&+& k~\delta_{n,N_{\beta}^{(i\pm1)}}~\delta_{j,i\pm1}~P_{j}(n;t,T),
\label{F}
\end{eqnarray}
where the '$+$' and '$-$' signs in the indices of the Kronecker delta 
are for $i = 1$ and $2$, respectively. 

One can have the following compact representation of the set of 
equations given by Eq.(\ref{F}) for all possible $n$ and $i$ values
\begin{equation}
\frac{\partial{{\bf P}(t,T)}}{\partial t} = {\bf A}(T){\bf P}(t,T),
\label{G}
\end{equation}
where $P_{1}(n;t,T)$ for $n = 0, 1, ..., {N_{\beta}^{(1)}}$ and 
$P_{2}(n;t,T)$ for $n = 0, 1, ..., {N_{\beta}^{(2)}}$ together comprise 
the elements of the column vector ${\bf P}(t,T)$ and ${\bf A}$ is the 
transition matrix of order $N = N_{\beta}^{(1)} +  N_{\beta}^{(2)} + 2$. If $G_{T}(i,t|j,0)$ be the Green's function that gives the probability 
to be in the state $i$ at a later time $t$ given that the system is in 
the state $j$ at time $t^{\prime} = 0$, the temperature being kept 
constant at temperature $T$, the matrix of Green's functions also 
satisfies the rate equation
\begin{equation}
\frac{d{\bf G}_{T}(t)}{dt} = {\bf A}(T){\bf G}_{T}(t)
\label{H}
\end{equation}
with the initial condition ${\bf G}_{T}(0) = {\bf I}$, where ${\bf I}$ 
is the identity matrix of order $N$. In terms of Green's functions, one 
can then rewrite the energy autocorrelation function as
\begin{equation}
<E(t,T)E(0,T)> =  \displaystyle \sum_{i = 1}^{N}
\displaystyle \sum_{j = 1}^{N} G_{T}(i,t|j,0)E_{i}E_{j}P_{eq}(j,T),    
\label{I}
\end{equation}
where $P_{eq}(j,T)$ is the equilibrium probability of the state $j$ at 
$T$. We write $\hat{G}_{T}(i,s|j)$ as the Laplace transform of 
$G_{T}(i,t|j,0)$:
\begin{equation}
\hat{G}_{T}(i,s|j) = \int\limits_{0}^{\infty} dt e^{-st} G_{T}(i,t|j,0). 
\label{J}
\end{equation}
The frequency dependent heat capacity is then given by
\begin{eqnarray}
C(\omega, T)&=& \frac{<E^2(T)>}{k_{B}T^2} \nonumber \\
&-& \frac{s}{k_{B}T^2} 
\displaystyle \sum_{i = 1}^{N} \displaystyle \sum_{j = 1}^{N} 
\hat{G}_{T}(i,s|j)E_{i}E_{j}P_{eq}(j,T). \nonumber \\
\label{K}
\end{eqnarray}
The computational procedure involves the numerical evaluation of the 
Green's functions by an inversion of matrix:
\begin{equation}
\hat{{\bf G}}_{T}(s) = (s{\bf I} - {\bf A}(T))^{-1},
\label{L}
\end{equation}
and that of ${\bf P_{eq}}(T)$ from the eigenvector corresponding 
to the zero eigenvalue of ${\bf A}(T)$. In the next section, we present
the results with discussion and note the relevance of our results.

The dynamical response of the system in the present framework is expected to
be determined by a set of parameters that includes the number $N_{\beta}$ of TLSs in a metabasin, the energy separation $\epsilon$ between the two levels
of a TLS, the energy of activation $\epsilon^{\ddagger}$ for barrier 
crossing within a TLS, the energy asymmetry $\Delta$ between the two 
adjacent metabasins, and
the critical number $N_{c}$ of TLSs required to be in the excited levels at 
a particular time for the $\alpha$-relaxation to occur. The choice of these 
parameters has been kept simple in this work though at the
expense of being {\it ad hoc} at least in some cases. For example, we have 
taken $N_{c}$ to be equal to $N_{\beta}$ and the value of $\epsilon$ has 
been taken of the order of $k_{B}T_{m}$, $T_{m}$ being the melting 
temperature. An approximate estimate of $N_{\beta}$ has been taken from 
simulations results \cite{hopping3, hopping4}. The value of the activation
energy is a rather difficult guess. We have used the guidance
provided by an earlier work of ours \cite{Chakrabarti-Bagchi-JCP}.
                                                                             
\noindent{\bf IV. RESULTS AND DISCUSSION}

\begin{figure}[tb]
\epsfig{file=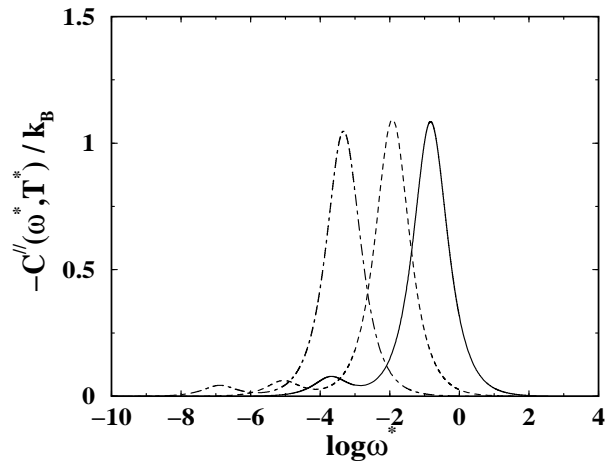,height=6cm,width=8cm,angle=0}
\caption{Frequency dependence of the imaginary 
part of the dynamic heat capacity 
$C^{\prime\prime}(\omega^{\star}, T^{\star})$ for our model system 
with $N_{\beta}^{(1)} = 3$ and 
$N_{\beta}^{(2)} = 5$, at three dimensionless reduced temperatures 
$T^{\star} = 0.9$ (solid line), $T^{\star} = 0.8$ (dashed line), and 
$T^{\star} = 0.7$ (dot-dashed line). Temperature $T$ is scaled by the 
melting temperature $T_{m}$ to have a reduced temperature 
$T^{\star}= T/T_{m}$. 
Frequency is also scaled by the inverse of $\tau_{1}(T_{m})$ to get a 
dimensionless reduced frequency $\omega^{\star} = \omega \tau_{1}(T_{m}) $. As in the earlier work,\cite{Chakrabarti-Bagchi-JCP} we set $k = 2.0$ in the reduced units, and $\epsilon = 2 k_{B}T_{m}$ and 
$\epsilon^{\ddagger}_{1} = 18k_{B}T_{m}$, the latter being the 
energy barrier to the transition from the level $1$ in a TLS. The 
presence of a fixed energy barrier to transition from one level to the 
another within a TLS is expected to impart an Arrhenius temperature 
dependence of the $\beta$-relaxation within the present framework. The 
laboratory glass transition temperature $T_{g}$ occurs at a temperature 
around $(2/3) T_{m}$.\cite{Angell-Ngai-McKenna-McMillan-Martin} All 
temperatures investigated here therefore fall between $T_{m}$ and  
$(2/3) T_{m}$. The same set of parameter values has been used for all
calculations in the present work unless it is specifically mentioned
otherwise.}
\end{figure}
In {\bf Figure 3}, we show the frequency dependence of the imaginary 
part of the heat capacity $C^{\prime\prime}(\omega^{\star},T^{\star})$ 
calculated for our model system at three different temperatures. The 
two-peak structure corresponding to the bimodal relaxation behavior
as conceived in the model is evident at all temperatures investigated. 
The peak at high frequencies {\it corresponds to the $\beta$-relaxation 
and is remarkably dominant}. The low-frequency peak is {\it due to the 
$\alpha$-relaxation} that occurs on a longer timescale. Note that 
{\it the separation between the position of the $\beta$-peak and that 
of the $\alpha$-peak grows as temperature is lowered.} This is 
indicative of the stringent constraint on the $\alpha$-process that 
arises from the cooperativity requirement. 

\begin{figure}[tb]
\epsfig{file=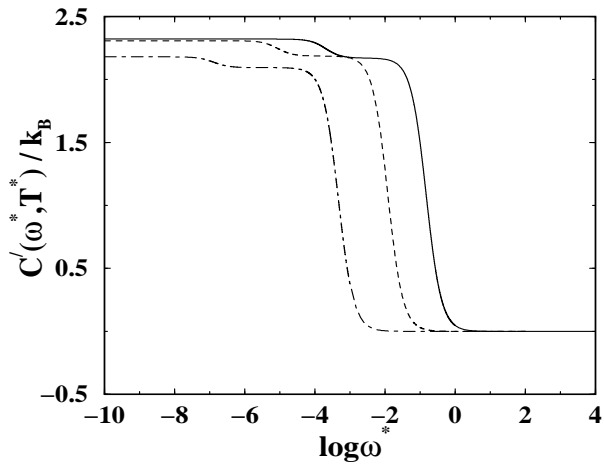,height=6cm,width=8cm,angle=0}
\caption{Frequency dependence of the real part of the 
dynamic heat capacity $C^{\prime}(\omega^{\star}, T^{\star})$ for our 
model system at three reduced temperatures $T^{\star} = 0.9$ 
(solid line), $T^{\star} = 0.8$ (dashed line), and $T^{\star} = 0.7$ 
(dot-dashed line).}
\end{figure}  
The features of the two-step relaxation also get reflected in the 
frequency spectrum of the real part of the heat capacity 
$C^{\prime}(\omega^{\star},T^{\star})$ as shown in {\bf Figure 4} at 
three different temperatures. This is expected on the basis of the 
Kramers-Kronig relation that relates the real and imaginary parts. 
Whenever $\omega^{-1}$ is on the order of the
timescale of a characteristic relaxation process, the system takes up 
energy inducing an increase in the real part of the heat capacity
around that frequency. The dominance of $\beta$-relaxation is again 
evident from a much larger increase at the high frequencies. The 
low-frequency limit corresponds to the static heat capacity of the 
system.

\begin{figure}[tb]
\epsfig{file=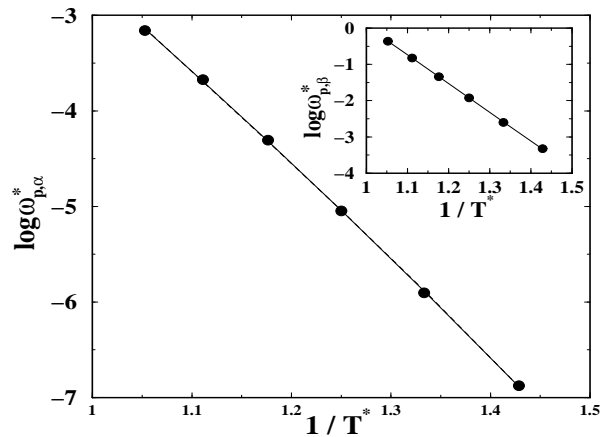,height=6cm,width=8cm,angle=0}
\caption{The $\alpha$-peak frequency 
$\omega_{p,\alpha}^{\star}$, on a logarithmic scale, versus 
the inverse temperature $1/T^{\star}$. The solid line and dashed line 
correspond to two nearly indistinguishable fits to the data with the 
Vogel-Fulcher-Tammann equation and a three-parameter scaling law (see text). From the fits, $T_{0,VFT} = 0.154$  and $T_{0,scl} = 0.475$ in the reduced 
units. The inset shows the temperature dependence of the $\beta$-peak 
frequency $\omega_{p,\beta}^{\star}$ in a $log \omega_{p,\beta}^{\star}$ versus the $1/T^{\star}$ plot. The solid line is a linear fit to the 
data with a slope of $7.912$ in the reduced temperature units.}
\end{figure} 
Let us now discuss the temperature dependence of the positions of the
peaks as they appear in the frequency spectrum of 
$C^{\prime\prime}(\omega^{\star},T^{\star})$. The temperature 
dependence of the $\alpha$-peak position 
$\omega_{p,\alpha}^{\star}(T^{\star})$ in the reduced scale is shown 
in {\bf Figure 5} in a $log~\omega_{p,\alpha}^{\star}(T^{\star})$ versus 
$1/T^{\star}$ plot. The data fit well to the Vogel-Fulcher-Tammann (VFT) 
equation:
$\omega_{p,\alpha}^{\star}(T^{\star}) = \omega_{0,VFT} exp[-A_{VFT}/
(T^{\star} - T_{0,VFT})]$. The small curvature of the fitted curve is, 
however, notable. This has been observed experimentally as 
well \cite{Birge-Nagel-PRL}. As in Ref.[4], we show in {\bf Figure 5} a 
second fit to the data with a scaling law: 
$\omega_{p,\alpha}^{\star}(T^{\star}) = \omega_{0,scl}[(T^{\star} - 
T_{0,scl})/T_{0,scl}]$. The two fits are nearly indistinguishable. The 
inset of {\bf Figure 5} shows the temperature dependence of the 
$\beta$-relaxation peak position $\omega_{p,\beta}^{\star}(T^{\star})$ 
in a similar plot. The linear fit corresponds to the Arrhenius behavior 
as expected from the supposition of the model.

\begin{figure}[tb]
\epsfig{file=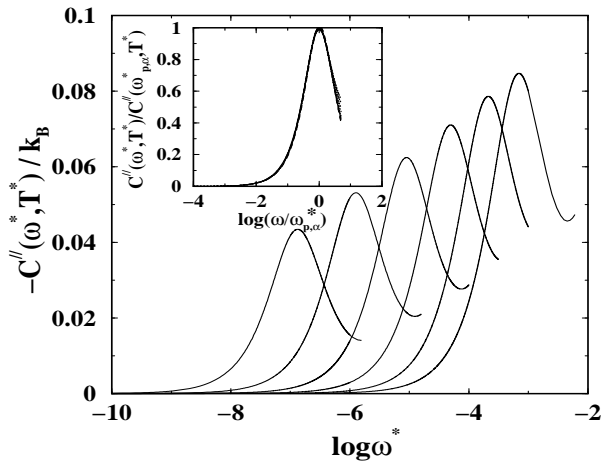,height=6cm,width=8cm,angle=0}
\caption{The frequency spectra of the imaginary part of
the dynamic heat capacity $C^{\prime\prime}(\omega^{\star}, T^{\star})$ 
at different temperatures showing the low-frequency $\alpha$-peak only. 
From the right to the left 
$T^{\star} = 0.95, 0.9, 0.85, 0.8, 0.75, 0.7$, respectively. The inset shows the same curves, in dotted lines, scaled by the height of the respective 
peaks versus $log(\omega^{\star}/\omega_{p,\alpha}^{\star})$. 
The solid line is a fit to the curve for $T^{\star} = 0.7$ with a 
three-parameter equation that is a good frequency domain representation 
of the time domain Kohlrausch-Williams-Watts stretched exponential form. The dashed line is a fit to the same curve with the Debye response 
function. The curves along with the two fits are nearly 
indistinguishable. The details of the fit are given in the text. The 
range of frequency shown here is restricted at the high frequency side 
due to the presence of the $\beta$-peak that leads to a break down of 
the fitting.}
\end{figure} 
Since the experiments till date could probe only the $\alpha$-relaxation
regime of the frequency spectrum, we have investigated the features
of the $\alpha$-relaxation peak as predicted by our model system in a 
bit more detail. In {\bf Figure 6}, we concentrate on the 
$\alpha$-relaxation regime of the frequency spectrum of 
$C^{\prime\prime}(\omega^{\star},T^{\star})$ at different temperatures 
to get an enlarged view of the $\alpha$-relaxation peak. It is evident 
that the area under the $\alpha$-peak becomes smaller as temperature 
drops. This implies that the cooperative component of the 
configurational part of the heat capacity diminishes with decreasing 
temperature. See Ref.[14] for a detailed discussion. Although the height 
and the position of the $\alpha$-peak have been found to change with 
temperature, its {\it shape} seems to show no such dependence. This
can be evident if one constructs a master plot by scaling both the 
height of the peak and the frequency. This is demonstrated in the 
inset of {\bf Figure 6} where we plot 
$C^{\prime\prime}(\omega^{\star},T^{\star})/C^{\prime\prime}
(\omega_{p,\alpha}^{\star},T^{\star})$ versus 
$log(\omega^{\star}/\omega_{p,\alpha}^{\star})$ at different 
temperatures for the frequency range of interest. The curves at 
different temperatures appear to be superimposed on each other with 
negligible error. It is therefore reasonable to conclude that the shape 
of the $\alpha$-peak here does remain independent of temperature. 
This is also in {\it remarkable agreement} with the observation in the 
computer simulation study reported in Ref.[14]. 

\begin{figure}[tb]
\epsfig{file=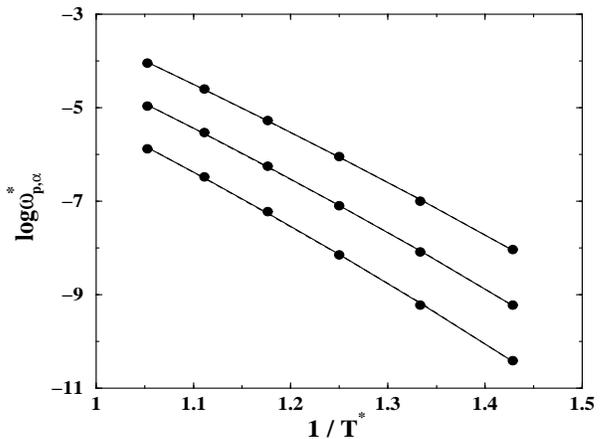,height=6cm,width=8cm,angle=0}
\caption{The $\alpha$-peak frequency 
$\omega_{p,\alpha}^{\star}$, on a logarithmic scale, versus 
the inverse temperature $1/T^{\star}$ for three different sets of
$\{N_{\beta}^{(1)}, N_{\beta}^{(2)}\}$ values. For each set, the solid 
line and dashed line correspond to two nearly indistinguishable fits to 
the data with the Vogel-Fulcher-Tammann equation and a three-parameter 
scaling law. From the fits, $T_{0,VFT} = 0.174$  and $T_{0,scl} = 0.485$ for $\{4, 6\}$ (top), $T_{0,VFT} = 0.197$  and $T_{0,scl} = 0.496$ for
$\{5, 7\}$ (middle), and $T_{0,VFT} = 0.203$  and $T_{0,scl} = 0.498$ 
for $\{6, 8\}$ (bottom).}
\end{figure} 
The results presented here till now have been for a single set of 
$N_{\beta}^{(1)}$, $N_{\beta}^{(2)}$, $\epsilon$ and
$\epsilon^{\ddagger}_{1}$ values. It is important to know how the 
predicted results are dependent on the choice of model parameters, in 
particular on $N_{\beta}^{(1)}$ and $N_{\beta}^{(2)}$, before one judges
the merits of the model. To this end, we have explored the parameter
dependence of the predictions of our model. The two-step relaxation 
behavior has been found to be revealed in the predicted heat capacity 
spectra of the model for a few other choices of 
$\{N_{\beta}^{(1)}, N_{\beta}^{(2)}\}$ as well, specifically, for 
$\{4,6\}$, $\{5,7\}$, and $\{6,8\}$. For each of these parameter sets,
a similar non-Arrhenius behavior of the temperature dependence of the 
$\alpha$-peak position has been observed as shown in {\bf Figure 7}. The 
Vogel temperature $T_{0,VFT}$ obtained from the VFT fit shifts to higher 
values along the series as one would expect from an increasingly 
stringent cooperativity requirement. Although the magnitude of the shift
is small, the trend is quite clear. The shape of the $\alpha$-peak has 
also been found to remain independent of temperature along the series
(data not shown). For $\{6,9\}$, the amplitude of the $\alpha$-peak 
gets diminished to a large extent by the time $T^{\star} = 0.7$ is 
reached, while the two-peak structure does not appear at all even at 
$T^{\star} = 0.95$ for $\{6,10\}$. The two-step relaxation behavior also
does not get evident from the predicted dynamic heat capacity 
spectra in the temperature range explored here for even $\{2,6\}$. It
follows therefore that a difference of $4$ in $N_{\beta}^{(2)}$ and 
$N_{\beta}^{(1)}$ with the set of $\epsilon$ and $\epsilon^{\ddagger}_{1}$ 
values used here is large enough to suppress any manifestation of the
two-step relaxation behavior within the present framework. On considering
varying asymmetry of the two adjacent metabasins, we further make
an interesting observation as illustrated in {\bf Figure 8}. Here, we have
varied $N_{\beta}^{(2)}$ with $N_{\beta}^{(1)}$, $\epsilon$ and
$\epsilon^{\ddagger}_{1}$ held fixed. For a symmetric
double well, only one peak appears in the frequency spectrum of the 
imaginary part of the dynamic heat capacity. The $\alpha$-peak disappears
as an $\alpha$-process is inconsequential from an energy consideration for
the symmetric double well. The amplitude of the $\alpha$-peak is the
largest with the least asymmetry and gradually diminishes with growing
asymmetry before it gets suppressed completely. The results suggest that 
the relative amplitudes of the $\alpha$-peak may provide an insight into 
the energy asymmetry between the metabasins in the energy landscape of the 
system.
\begin{figure}[tb]
\epsfig{file=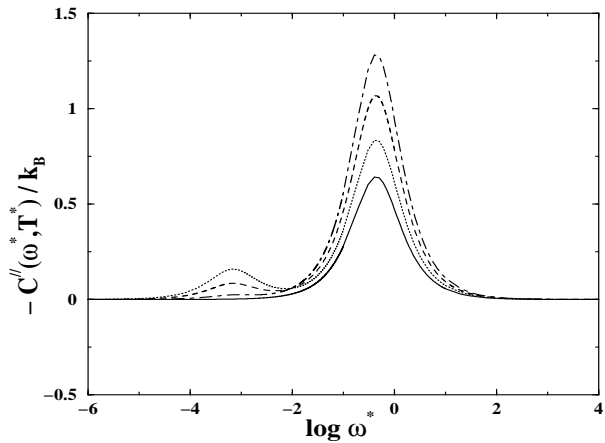,height=6cm,width=8cm,angle=0}
\caption{The frequency spectra of the imaginary part of
the dynamic heat capacity $C^{\prime\prime}(\omega^{\star}, T^{\star})$ 
at temperature $T^{\star} = 0.95$ for four choices of $N_{\beta}^{(2)}$
with $N_{\beta}^{(1)}$ kept fixed: (a) $\{3, 3\}$ (solid line), 
(b) $\{3, 4\}$ (dotted line), (c) $\{3, 5\}$ (dashed line), and
(d) $\{3, 6\}$ (dot-dashed line).}
\end{figure}  

In the inset of {\bf Figure 6}, we have also shown a fit to the curve for 
$T^{\star} = 0.7$ with a three-parameter equation that is a good frequency 
domain representation of the time domain Kohlrausch-Williams-Watts (KWW) 
stretched exponential form \cite{Bergman-JAP-2000}. From the fit 
parameters we obtain the stretching parameter $\beta_{KWW} = 0.94$. Such a 
$\beta_{KWW}$ value implies very weakly nonexponential behavior. In fact, a 
fit with the Debye response function that corresponds to a single 
exponential behavior is found to be reasonable and for some of the other 
sets of $\{N_{\beta}^{(1)}, N_{\beta}^{(2)}\}$, is found to be as good as 
the one with the three-parameter equation corresponding to a stretching 
parameter very close to unity. This is reasonable as spatially heterogeneous domains, which is believed to be the primary reason for the stretched 
exponential relaxation in supercooled 
liquids \cite{Ediger-ARPC, Sillescu-JNCS, Richert-JPCM}, has not been 
considered in the present calculations. The heterogeneous dynamics in 
different domains can be included either through a distribution of 
$\epsilon$ (the separation between the energy levels within
a TLS) or through a distribution of barrier height for transition from one 
level to the other within a TLS. When the heterogeneity is included, the 
exponent $\beta_{KWW}$ is expected to decrease considerably as indeed found 
in the treatment of structural relaxation within a similar model where we 
have considered a distribution of $\epsilon$ 
\cite{Chakrabarti-Bagchi-unpublished}.    

The remarkably dominant $\beta$-peak as predicted by our model merits 
further consideration. In order to trace back its origin, we note that the 
constraint of cooperativity on the $\alpha$-relaxation allows the system to 
take up energy only through the localized $\beta$-processes unless the 
condition is satisfied. The system takes up more energy through the 
excitation of individual $\beta$-processes (that collectively bring about 
the $\alpha$-process) than the $\alpha$-process itself which corresponds to 
a transition from one metabasin to another. This argument is substantiated
in {\bf Figure 9}, where we focus on the amplitude of the $\beta$-peak 
relative to that of the $\alpha$-peak in the frequency spectrum of the 
imaginary part of the dynamic heat capacity. Note that the $\alpha$-peak is 
not distinctly observed for $\{ 1, 3\}$ as the separation between the two 
peaks is not enough. A decrease in the requirement of cooperativity for the 
$\alpha$-process is found to result in a fall in the relative amplitude of 
the $\beta$-peak. One can, therefore, define a parameter 
$Q_{hc}^{\alpha\beta}$ as the ratio of the amplitude of 
the $\beta$-peak to that of the $\alpha$-peak, which is in general 
temperature dependent. That is, $Q_{hc}^{\alpha\beta}(T)= C^{\prime\prime}(\omega_{p,\beta},T)/C^{\prime\prime}(\omega_{p,\alpha},T)$. The parameter 
$Q_{hc}^{\alpha\beta}(T)$ may serve as a measure of the cooperativity needed for $\alpha$-relaxation to take place.   
\begin{figure}[tb]
\epsfig{file=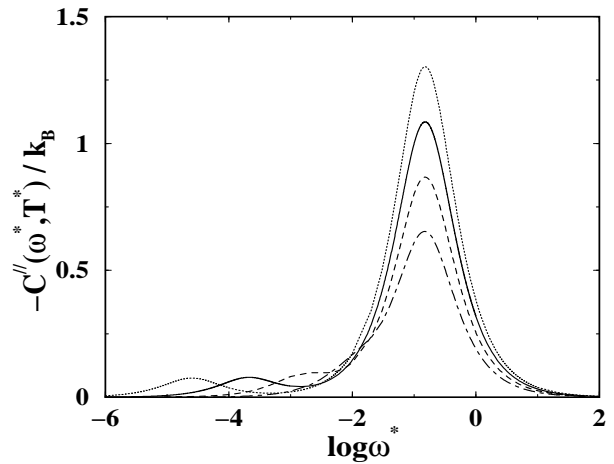,height=6cm,width=8cm,angle=0}
\caption{The frequency spectra of the imaginary part of
the dynamic heat capacity $C^{\prime\prime}(\omega^{\star}, T^{\star})$ 
at temperature $T^{\star} = 0.9$ for four sets of 
$\{N_{\beta}^{(1)}, N_{\beta}^{(2)}\}$ values with 
$N_{\beta}^{(2)}- N_{\beta}^{(1)}$ held fixed:
(a) $\{4, 6\}$ (dotted line), (b) $\{3, 5\}$ (solid line),
(c) $\{2, 4\}$ (dotted line), and
(d) $\{1, 3\}$ (dot-dashed line).}
\end{figure} 

Let us finally consider a limiting case where the $\beta$-processes are not
associated with any energy cost, {\it i.e.} $\epsilon = 0$. We then describe a $\beta$-process as an activated event in a two-state system (rather then
a two-level system) with a nonzero energy of activation 
$\epsilon^{\ddagger}$ for barrier crossing. We further assume a nonzero 
value of $\Delta$ and one of the $k_{i}$'s to be temperature independent, 
where $k_{i}$ is the rate of transition from the metabasin $i$ (See 
{\bf Figure 2}). In particular, we assume $k_{2}$ to be independent of 
temperature while $k_{1} = k_{2}~exp[-\Delta /(k_{B}T)]$. From the results
we have already presented here, one would expect then the imaginary part of 
the dynamic heat capacity to show only one peak. This has indeed been 
observed as shown in {\bf Figure 10}.    
\begin{figure}[tb]
\epsfig{file=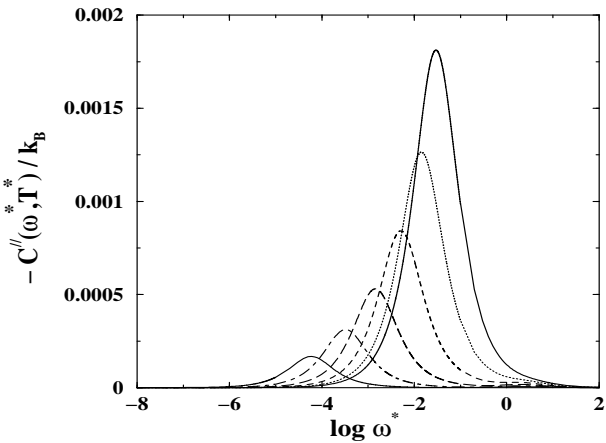,height=6cm,width=8cm,angle=0}
\caption{The frequency spectra of the imaginary part of
the heat capacity $C^{\prime\prime}(\omega^{\star}, T^{\star})$ at 
different temperatures for the limiting case considered with 
$N_{\beta}^{(1)} = 3$, $N_{\beta}^{(2)} = 5$, $\epsilon = 0$,  
$\epsilon^{\ddagger} = 18k_{B}T_{m}$, and $\Delta = 4k_{B}T_{m}$. We
set $k_{2} = 2.0$ in the reduced units and the temperature dependence is
kept contained in $k_{1}$. From the right to the left 
$T^{\star} = 0.95, 0.9, 0.85, 0.8, 0.75, 0.7$, respectively.}
\end{figure} 
 
\noindent{\bf V. CONCLUSION}
         
Let us first summarize the present work. We have employed a kinetic model of glassy dynamics that considers cooperativity through the constraint of 
{\it $\beta$-organized-$\alpha$-process} within the landscape paradigm. The 
two-step relaxation behavior as conceived in the model gets revealed in the 
frequency dependent heat capacity of the model for a reasonable range of 
parameter values. The analysis of the predicted dynamic heat capacity 
spectra suggests the following emergent features of our model: 
({\it i}) The $\alpha$-peak frequency has a non-Arrhenius temperature 
dependence. ({\it ii}) The shape of the $\alpha$-peak as it appears in the 
frequency spectrum of the imaginary part of the dynamic heat capacity is 
invariant in temperature. ({\it iii}) The amplitude of the 
$\beta$-peak is considerably larger than that of the $\alpha$-peak.
({\it iv}) In a limiting case where the $\beta$-process involves an 
activated transitions between two states of the same energy while the two 
adjacent metabasins have an asymmetry in energy, the frequency spectrum of 
the dynamic heat capacity exhibits only a one-peak (one-step) structure.

While the first (the non-Arrhenius temperature dependence) of this lot 
conforms to a number of experimental results, the second one has been 
observed in a recent computer simulation study \cite{Scheidler-PRB-2001}. 
The third one, however, does invite a few comments. Although a remarkably 
dominant high-frequency peak has been reported in Ref.[14], it is rather 
ascribed to the vibrational excitations of the system. We, however, note 
that {\it the predominance of the $\beta$-peak is not in contention of 
being contradicted with any published experimental data on the dynamic heat 
capacity.} In this context, it may be reasserted that the $\beta$-relaxation regime is still beyond the scope of the specific heat spectroscopy since the existing experimental setup can scan a frequency range up to only $10^4$ Hz. Nevertheless, such a dominant $\beta$-peak is unlikely to be observed in 
typical fragile liquids where even for a deeply supercooled state, the 
signature of the $\beta$-relaxation in the frequency spectrum of the dynamic structure factor is only a weak secondary peak at higher frequencies. 
 
The present work suggests that the relative amplitudes of the two peaks may 
provide insight into the microscopic mechanism of the relaxation processes
in deeply supercooled liquids. In view of this, further measurements of 
frequency dependent specific heat of various glass forming liquids will 
certainly be worthwhile. In a recent simulation study of computer model of 
amorphous $Ni_{81}B_{9}$, van Ee {\it et. al.} have shown that the hopping 
mode is not only collective but appears to involve rather large number of 
neighbors \cite{van EE-PRE-1998}. This study seems to support the picture 
that vibrational coherence among neighbors is a prerequisite for large 
scale hopping. In such a scenario, our $\beta$-process may be identified 
with such vibrational motion while hopping is the jump between the two 
minima shown in {\bf Figure 2}. We further argue that a predominant 
$\beta$-peak would be a characteristic feature of Stillinger's picture 
which assumes an $\alpha$-process to involve a concerted series of 
$\beta$-processes \cite{Wales-2003}. However, proper characterization of the
$\beta$-processes remains unclear. It may be noted here in this respect that
the activated dynamics with an Arrhenius type temperature dependence 
characterizing the $\beta$-process in the present work is the typical 
feature of what is called the 'slow' $\beta$-process \cite{Wales-2003}. 
However, the manifestation of this dynamical process with a predominant peak
in the frequency spectrum of the dynamic heat capacity is more likely to be 
relevant with an even faster $\beta$-process typically appearing in the 
mode-coupling theory predictions.
   
Several other comments on the present work are in order. First, the present 
model can be taken to belong to the class of kinetically constrained 
models that attempts to provide a description of glassy dynamics by imposing dynamical constraints on the allowed transitions between different 
configurations of the system, while maintaining the detailed balance
\cite{Ritort-Sollich-AP}. In particular, our model resembles the 
models of hierarchically constrained dynamics of glassy relaxation, due 
originally to Palmer {\it et. al.} \cite{Palmer-PRL-1984}, in the spirit 
that brings in cooperativity. Second, the high-frequency peak for real 
liquids is likely to draw contribution from $\beta$-relaxation as well as 
vibrational excitations. In some cases, specifically at low temperatures 
close to the glass transition, it is possible that these two can be 
sufficiently separated to give rise to an additional peak in the imaginary 
part of the frequency dependent specific heat. Third, it is imperative to 
compare the two-peak structure of the frequency spectrum of the imaginary 
part of the dynamic heat capacity with its dielectric analog. In contrary to the prediction our model makes on dynamic heat capacity behavior, it is 
rather the predominance of the $\alpha$-peak that has been observed in a 
vast body of experimental data on dielectric relaxation. While the 
fluctuation in energy within the present framework of the model that has an 
energy landscape picture at the backdrop translates easily into the 
calculation of the frequency dependent heat capacity, the model as such does not allow us to calculate the frequency dependence of the dielectric 
constant. The latter needs further development of the model. However, the 
well-known bimodal frequency dependence of the dielectric relaxation in 
supercooled liquids can be at least qualitatively understood from the 
present description of $\beta$- and $\alpha$-processes. We essentially 
follow the description of Lauritzen and Zwanzig in assuming that a 
$\beta$-process can be taken to correspond to a two-site angular jump of 
individual molecules by a small angle around some axis 
\cite{Lauritzen-Zwanzig}. These individual, uncorrelated angular jumps lead 
to a partial relaxation of the total electric moment ${\bf M}(t)$ of the 
whole system (note that ${\bf M}(t)$ is the sum of the dipole moment of the 
individual molecules). The dielectric susceptibility spectrum can be 
obtained from the auto-time correlation function of ${\bf M}(t)$ by using 
the linear response theory \cite{Zwanzig-book}. Since ${\bf M}(t)$ is a sum 
of a relatively large number of individual dipole moments, the former is a 
Gaussian Markov process and thus the time correlation function of the 
$\beta$-relaxation mediated part must decay exponentially. As noted earlier, this $\beta$-relaxation mediated decay is incomplete because all the jumps 
are small and restricted. Thus, it is fair to assume the following form for 
the auto-time correlation function of ${\bf M}(t)$
\begin{eqnarray}
C_{M}(t) &=& <M_{\beta}^{2}>~exp(-t/\tau_{\beta}) \nonumber \\
&+& (<M_{\circ}^{2}> - <M_{\beta}^{2}>)~exp(-t/\tau_{\alpha}),
\label{DR1}
\end{eqnarray}
where $\tau_{\beta}$ and $\tau_{\alpha}$ are the time scales of 
$\beta$-relaxation and $\alpha$-relaxation, respectively. In the above 
equation $ <M_{\beta}^{2}>$ is the value by which the mean-square total 
dipole moment decays due to $\beta$ relaxation alone from the initial value 
of $<M_{\circ}^{2}>$. The rest of the decay ({\it i. e.} from 
$<M_{\circ}^{2}> - <M_{\beta}^{2}>$) to zero occurs via the 
$\alpha$-process. This suggests that with well separated $\tau_{\beta}$ and 
$\tau_{\alpha}$, one would observe bimodal dispersion. However, the 
calculation of $ <M_{\beta}^{2}>$ would require a more detailed model than 
the one attempted here. 

The present study suggests several future problems. First, it would be 
interesting to investigate the frequency dependence of the specific heat of 
several molecular liquids of varying fragility in computer simulations. In 
some cases, one should be able to discern a three-peak structure. Second, it
would be also of interest to investigate if the frequency dependent specific
heat can be used to study the so called Boson peak which has drawn much 
attention in recent times \cite{Lubchenko-Wolynes-PNAS}. Third, a comparison 
of the frequency dependence of specific heat with that of the dynamic 
structure factor of certain model systems would provide us insight to 
decide on whether the microscopic mechanisms of relaxation vary for 
different modes. Finally, a generalization of the present model or an 
altogether different model to describe the frequency dependence of dynamic 
heat capacity and that of dielectric response in the same framework would be a worthwhile undertaking.

\noindent{\bf ACKNOWLEDGMENTS}

We thank Professor N. Menon and Professor C. A. Angell for useful 
discussions. This work was supported in parts by grants from CSIR and DST, 
India. DC acknowledges the University Grants Commission (UGC), India for 
providing the Research Fellowship.


\begin{references}

\bibitem{Angell-Ngai-McKenna-McMillan-Martin} C. A. Angell, K. L. Ngai, 
G. B. McKenna, P. F. McMillan and S. W. Martin, J. Appl. Phys. 
{\bf 88}, 3113 (2000).

\bibitem{Debenedetti-Stillinger-Nature-2001} P. G. Debenedetti and F. H. 
Stillinger, Nature {\bf 410}, 259 (2001).

\bibitem{Mohanty-ACP-1994} U. Mohanty, Adv. Chem. Phys. {\bf 89}, 89 
(1994).

\bibitem{Birge-Nagel-PRL} N. O. Birge and S. R. Nagel, Phys. Rev. Lett. 
{\bf 54}, 2674 (1985).

\bibitem{Christensen} T. Christensen, J. Phys. (Paris) Colloq. {\bf 46}, 
C8-635 (1985).

\bibitem{Zwanzig-JCP-1988} R. Zwanzig, J. Chem. Phys. {\bf 88}, 5831 
(1988).

\bibitem{Birge-PRB-1986} N. O. Birge, Phys. Rev. B {\bf 34}, 1631 (1986).

\bibitem{Dixon-Nagel-PRL-1988} P. K. Dixon and S. R. Nagel, Phys. Rev. 
Lett. {\bf 61}, 341 (1988).

\bibitem{Menon-JCP-1997} N. Menon, J. Chem. Phys. {\bf 105}, 5246 
(1996). 

\bibitem{Oxtoby-JCP-1986} D. W. Oxtoby, J. Chem. Phys. {\bf 85}, 1549 
(1986).

\bibitem{Jackle-PhysicaA} J. J\"{a}ckle, Physica A {\bf 162}, 377 
(1990).

\bibitem{Nielsen-Dyre-PRB} J. K. Nielsen and J. C. Dyre, Phys. Rev. B 
{\bf 54}, 15754 (1996). 

\bibitem{Nielsen-PRE-1999} J. K. Nielsen, Phys. Rev. E {\bf 60}, 471 
(1999).

\bibitem{Scheidler-PRB-2001} P. Sceidler, W. Kob, A. Latz, J. Horbach,
and K. Binder, Phys. Rev. B {\bf 63}, 104204 (2001).

\bibitem{Chakrabarti-Bagchi-JCP} D. Chakrabarti and B. Bagchi, J. Chem.
Phys. {\bf 120}, 11678 (2004).

\bibitem{Lebowitz-Percus-Verlet} J. L. Lebowitz, J. K. Percus, and L. 
Verlet, Phys. Rev. {\bf 153}, 250 (1967).

\bibitem{Harbola-Das-PRE} U. Harbola and S. P. Das, Phys. Rev. E 
{\bf 64}, 46122 (2001).

\bibitem{Goldstein-JCP-1969} M. Goldstein, J. Chem. Phys. {\bf 51}, 3728
(1969).

\bibitem{Johari-Goldstein} G. P. Johari and M. Goldstein, J. Chem. Phys. 
{\bf 53}, 2372 (1970); {\it ibid.} {\bf 55}, 4245 (1971).

\bibitem{Stillinger-Science-1995} F. H. Stillinger, Science {\bf 267}, 
1935 (1995).

\bibitem{Sastry-Nature-1998} S. Sastry, P. G. Debenedetti, and 
F. H. Stillinger, Nature {\bf 393}, 554 (1998).

\bibitem{Bengtzelius} U. Bengtzelius, W. G\"{o}tze, and A. 
Sj\"{o}lander, J. Phys. C {\bf 17}, 5915 (1984).

\bibitem{Gotze-Sjogren} W. G\"{o}tze and L. Sj\"{o}gren, Rep. Prog. 
Phys. {\bf 55}, 241 (1992).

\bibitem{Angell-JPCS} C. A. Angell, J. Phys. Chem. Solids {\bf 49}, 863 
(1988).

\bibitem{Bagchi-JCP-1994} B. Bagchi, J. Chem. Phys. {\bf 101}, 9946 (1994).

\bibitem{hopping1} G. Wahnstr\"{o}m, Phys. Rev. A {\bf 44}, 3752 (1991).

\bibitem{hopping2} H. Miyagawa, Y. Hiwatari, B. Bernu, and J. P. Hansen, 
J. Chem. Phys. {\bf 88}, 3879 (1988).

\bibitem{hopping3} S. Bhattacharyya and B. Bagchi, Phys. Rev. Lett, 
{\bf 89}, 25504 (2002).

\bibitem{hopping4} S. Bhattacharyya, A. Mukherjee, and B. Bagchi, J. Chem. 
Phys, {\bf 117}, 2741 (2002).

\bibitem{Donati-et-al-PRL-1998} C. Donati {\it et. al.}, Phys. Rev. Lett. 
{\bf 80}, 2338 (1998).

\bibitem{Kim-Keyes-JCP-2004} J. Kim and T. Keyes, J. Chem. Phys. {\bf 121},
4237 (2004).
 
\bibitem{Dyre-PRE-1999} J. C. Dyre, Phys. Rev. E {\bf 59}, 2458 (1999).

\bibitem{van Kampen} N. G. van Kampen, {\it Stochastic Processes in 
Physics and Chemistry} (Elsevier Science, Amsterdam, 1992).

\bibitem{Bergman-JAP-2000} The three-parameter equation in the
frequency domain to fit a $X^{\prime\prime}(\omega)$ versus $\omega$ 
curve reads as
$X^{\prime\prime}(\omega) = X_{p}^{\prime\prime}(\omega)/[1-b+b/(1+b)
\{b(\omega_{p}/\omega)+(\omega/\omega_{p})^b\}]$, where 
$X_{p}^{\prime\prime}$, $\omega_{p}$, and $b$ are the peak height, peak
position, and the shape parameter, respectively. These give the
KWW parameters through definite relationships. In particular,
$b \approx \beta_{KWW}$. See R. Bergman, J. Appl. Phys. {\bf 88}, 1356 
(2000).

\bibitem{Ediger-ARPC}  M. D. Ediger, Annu. Rev. Phys. Chem. {\bf 51}, 99 
(2000).

\bibitem{Sillescu-JNCS} H. Sillescu, J. Non-Cryst. Solids {\bf 243}, 81 
(1999).

\bibitem{Richert-JPCM} R. Richert, J. Phys.: Condens. Matter {\bf 14}, 
R703 (2002).

\bibitem{Chakrabarti-Bagchi-unpublished} D. Chakrabarti and B. Bagchi
(unpublished).   

\bibitem{van EE-PRE-1998} L. D. van Ee, B. J. Thijsee, and J. Sietsma, 
Phys. Rev. E {\bf 57}, 906 (1998).

\bibitem{Wales-2003} D. J. Wales, {\it Energy Landscapes} (Cambridge 
University Press, Cambridge, 2003).

\bibitem{Ritort-Sollich-AP} F. Ritort and P. Sollich, Adv. Phys. 
{\bf 52}, 219 (2003).

\bibitem{Palmer-PRL-1984} R. G. Palmer, D. L. Stein, E. Abrahams, and
P. W. Anderson, Phys. Rev. Lett. {\bf 53}, 958 (1984). 

\bibitem{Lauritzen-Zwanzig} J. I. Lauritzen, Jr. and R. Zwanzig, Adv. Mol. 
Relax. Processes {\bf 5}, 339 (1973).

\bibitem{Zwanzig-book} R. Zwanzig, {\it Nonequilibrium Statistical 
Mechanics} (Oxford University Press, New York, 2001)

\bibitem{Lubchenko-Wolynes-PNAS} V. Lubchenko and P. G. Wolynes, Proc. Natl. Acad. Sci. USA, {\bf 100}, 1515 (2003).

\end{references}
\end{document}